\documentclass[%
 reprint, superscriptaddress,
 amsmath,amssymb,
 aps,
]{revtex4-2}

\usepackage{graphicx}% Include figure files
\usepackage{dcolumn}% Align table columns on decimal point
\usepackage{bm}% bold math
\usepackage{xcolor}
\usepackage[version=3]{mhchem}
\usepackage{gensymb}
\usepackage{textgreek}
\begin{document}

\preprint{APS/123-QED}

\title{Cell spheroid viscoelasticity is deformation-dependent}% Force line breaks with \\

\author{Ruben C. Boot}
\affiliation{
 Department of Chemical Engineering, Delft University of Technology, 2629 HZ Delft, The Netherlands
}
\author{Anouk van der Net}
\affiliation{
 Department of Bionanoscience, Kavli Institute of Nanoscience, Delft University of Technology, 2629 HZ Delft, The Netherlands
}
\author{Christos Gogou}
\affiliation{
 Department of Bionanoscience, Kavli Institute of Nanoscience, Delft University of Technology, 2629 HZ Delft, The Netherlands
}
\author{Pranav Mehta}
\affiliation{
 Department of Chemical Engineering, Delft University of Technology, 2629 HZ Delft, The Netherlands
 }
\affiliation{
 Department of Cell and Chemical Biology and Oncode Institute, Leiden University Medical Center, Leiden, The Netherlands
}
\author{Dimphna H. Meijer}
\affiliation{
 Department of Bionanoscience, Kavli Institute of Nanoscience, Delft University of Technology, 2629 HZ Delft, The Netherlands
}
\author{Gijsje H. Koenderink}
\affiliation{
 Department of Bionanoscience, Kavli Institute of Nanoscience, Delft University of Technology, 2629 HZ Delft, The Netherlands
}
\author{Pouyan E. Boukany}
 \email{p.e.boukany@tudelft.nl}
\affiliation{
 Department of Chemical Engineering, Delft University of Technology, 2629 HZ Delft, The Netherlands
}

\begin{abstract}
Tissue surface tension influences cell sorting and tissue fusion. Earlier mechanical studies suggest that multicellular spheroids actively reinforce their surface tension with applied force. Here we study this open question through high-throughput microfluidic micropipette aspiration measurements on cell spheroids to identify the role of force duration and cell contractility. We find that larger spheroid deformations lead to faster cellular retraction once the pressure is released, regardless of the applied force and cellular contractility. These new insights demonstrate that spheroid viscoelasticity is deformation-dependent and challenge whether surface tension truly reinforces.
\end{abstract}

\maketitle

%\tableofcontents

The physical response of multicellular tissues to an applied stress is critical in the regulation of various physiological processes, such as embryonic morphogenesis \cite{Hahn2009,Mammoto2010}, wound healing \cite{Brugues2014}, cell differentiation \cite{Discher2009}, and cancer metastasis \cite{Nia2016,Nia2020}. While the mechanical response of single cells depends on their cytoskeleton, plasma membrane and nuclear stiffness \cite{Galie2022,Friedl2011}, overall tissue mechanics is additionally dependent on intercellular adhesions and the extracellular environment \cite{Schiele2015,Heisenberg2013,Han2020}. When tissues form and merge, their resulting morphology is defined by this mechanical interplay between cells across multiple length scales, called tissue fluidity \cite{Jakab2008,Kosheleva2020,Grosser2021}.

To examine the relation between cellular mechanics and tissue fluidity, dissociated cells can be manipulated into a spherical assembly, termed spheroid, by letting them sediment and aggregate in a confined space. Spheroids have become a popular \textit{in vitro} model as they recreate both the multicellularity and three-dimensional (3D) microenvironment of \textit{in vivo} tissues \cite{Gonzalez-Rodriguez2012, Boot2021}. They round up over time to minimize surface energy, similar to liquid droplets \cite{Foty1996}. Previous studies have determined the apparent spheroid surface tension $\gamma$, which has been related to tissue spreading \cite{Ryan2001} and cell sorting \cite{Schotz2008}. Here, the magnitude of $\gamma$ and the related cellular arrangement depend on the interplay between the intercellular adhesion and cortical tension of the cells \cite{Foty2005,Brodland2002,Manning2010}. 

A conventional biophysical tool to measure $\gamma$ is micropipette aspiration (MPA), where the spheroid is partly aspirated into a micron-sized pipette under a constant stress and the advancing creep length $L(t)$ of the spheroid protrusion is monitored over time \cite{Dufour2010}. Aspirated spheroids behave like a viscoelastic liquid, first displaying an elastic response followed by an apparently linear viscous response. The constant creep rate $\dot{L_\infty}$ of the linear viscous response during both aspiration and relaxation of the spheroid (once the aspiration pressure is released) is used to calculate the spheroid surface tension $\gamma$ \cite{Dufour2010, Yousafzai2022}. Intriguingly, $\gamma$ was shown to depend on the applied aspiration pressure $\Delta P$, suggesting a reinforcement of $\gamma$ through an active response of the cells to the mechanical force \cite{Dufour2010}. However, no dependency of $\gamma$ on the applied force was observed in parallel-plate uniaxial compression experiments \cite{Foty1996,Schotz2008}. This raises the question whether cells actively reinforce their surface tension with the applied force or if the current understanding of viscoelastic spheroid aspiration data is insufficient.

\begin{figure}[b]
\includegraphics[width=8.5cm]{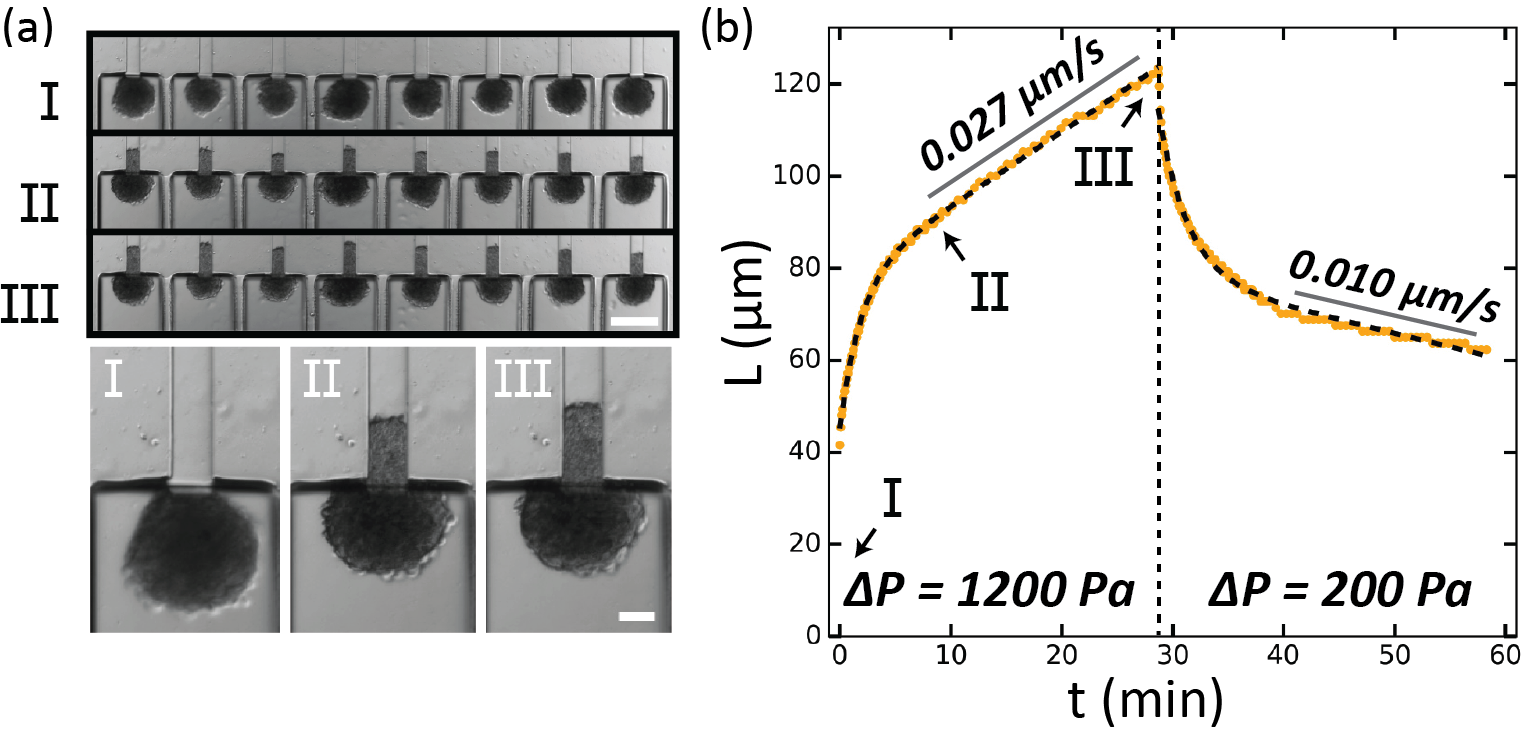}
\caption{\label{Fig1} Viscoelastic deformation of NIH3T3 spheroids. (a) Brightfield images of 8 NIH3T3 spheroids aspirated at 0 min (I), 10 min (II) and 30 min (III), with an overview of the microfluidic chip (top, scale bar 200 \textmu m), and a single spheroid close-up (bottom, scale bar 50 \textmu m). (b) The creep length $L (t)$, for the spheroid shown in (a, bottom), aspirated at 1200 Pa and left to retract at 200 Pa. The data (orange dots) is fitted with the Modified Maxwell model (black dashed lines), and the derived viscous creep rate values are added to the plot.}
\end{figure}

In this Letter, we address this question by studying how the duration that cells are exposed to different aspiration forces, alongside differences in cell mechanics, affect the tissue surface tension and its possible reinforcement. Recently, we have developed a microfluidic micropipette aspiration device that allows for a higher throughput than the conventional glass micropipette technique \cite{Boot2022}. Where the traditional technique only aspirates one spheroid at a time, our device can aspirate up to eight spheroids simultaneously [Fig. \ref{Fig1}(a)]. By flowing spheroids into individual parallel aspiration pockets that are aligned with squared constriction channels of 50x50 \textmu m\textsuperscript{2} (mimicking micropipettes), parallel creep tests can be applied by lowering a water reservoir attached to the outlet of the device [Supplemental Material and Supplementary Fig. 1 \cite{SupMat}]. First, a spheroid aspiration measurement is conducted, where the creep length $L(t)$ increases over time. Next, the pressure is released, thus starting a stress relaxation test, where the protrusion retracts over time.
 
By fitting the creep data with a modified Maxwell model [\cite{Dufour2010} and Supplemental Material \cite{SupMat}], the fast elastic deformation $\delta$ at short times and viscous flow with constant velocity $\dot{L_\infty}$ at long times can be defined for both the aspiration and retraction curves. Assuming volume conservation of the non-aspirated part of the spheroid, the aspiration force for a cylindrical pipette is given by $f = \pi R_{p}^{2} (\Delta P - \Delta P_{c})$, with $R_p$ the radius of the pipette and $\Delta P_c$ the critical pressure above which aspiration occurs \cite{Dufour2010}. Assuming that the viscosity $\eta$ of the spheroid remains unchanged during the aspiration and retraction phase, the critical pressure is deduced from $\Delta P_c = \Delta P \dot{L_\infty^r}/(\dot{L_\infty^r}+\dot{L_\infty^a})$, where $\dot{L_\infty^r}$ and $\dot{L_\infty^a}$ are the retraction and aspiration flow rates, respectively [for full derivation see Supplemental Material \cite{SupMat}]. By applying the Laplace law, the spheroid surface tension is derived from the critical pressure via $\Delta P_c =2\gamma(\frac{1}{R_p}-\frac{1}{R})$, with $R$ being the spheroid radius, which can be approximated by the initial radius $R_0$, as $R_p \ll R_0$. Following previous work by Davidson et al. \cite{Davidson2019,Son2007}, the effective channel radius $R_{eff}$ for our squared 50x50 \textmu m\textsuperscript{2} channel is 27 \textmu m [for derivation see Supplemental Material \cite{SupMat}].

Homogeneous spheroids of NIH3T3 fibroblast cells and human embryonic kidney (HEK293T) cells were formed using the Sphericalplate 5D (Kugelmeiers) and ranged between 65 and 125 \textmu m in radius through all experiments [Supplemental Material \cite{SupMat} and Supplementary Fig. 2]. All details on the microfluidic device can be found in our previous study \cite{Boot2022} and the Supplemental Material \cite{SupMat}. Only spheroids with a constant volume during aspiration were analyzed. During the stress relaxation test, bringing $\Delta P$ entirely back to zero often made spheroids move out of the pockets, preventing the monitoring of the protrusion retraction. This was likely due to the presence of a minor backflow in the microfluidic device, as manually bringing back the outlet reservoir to the exact same starting height proved to be difficult. To circumvent this, all retraction measurements were performed by leaving a minor pressure that still allowed for the protrusion to retract while keeping it in the constriction channel. This led to a small readjustment in the derivation of $\Delta P_c$ and $\gamma$ [Supplemental Material \cite{SupMat}].

First, creep aspiration tests were performed on NIH3T3 spheroids for 30 minutes (min), long enough for the protrusions to have entered the linear viscous regime (as the creep rate did not change anymore after 10 min). Next, stress relaxation tests were captured for an identical 30 min [Fig. \ref{Fig1}(b)]. We found that the derived viscous retraction flow velocity $\dot{L_\infty^r}$ was strongly influenced by the duration of retraction [Supplementary Fig. 3 \cite{SupMat}]. Even after a 2 hour-long retraction measurement, $\dot{L_\infty^r}$ still decreased over time as the creep curve plateaued [Supplementary Fig. 4 \cite{SupMat}]. Interestingly, in traditional MPA studies on murine sarcoma (S180) cell spheroids, where no remaining pressure was left during retraction, the flow velocity did appear to be linear over time \cite{Dufour2010,Yousafzai2022}. We hypothesize that the minor pressure left in our retraction measurements induced the plateau, which would mean that here retraction is not governed by one constant critical pressure $\Delta P_c$. Instead, the spheroid protrusion first retracts with a large $\Delta P_c$ upon release of the aspiration pressure, after which the creep curve plateaus due to the remaining pressure counteracting the spheroid now retracting with a smaller $\Delta P_c$. To eliminate active contraction during spheroid retraction, we treated the NIH3T3 spheroids with the myosin II inhibitor Blebbistatin and monitored their retraction at 200 Pa. Now, the spheroids first displayed a minor elastic retraction after which they started aspirating again [Supplemental Material and Supplementary Fig. 5 \cite{SupMat}]. This demonstrates how for these measurements retraction can not be governed by a constant $\Delta P_c$, as retraction changed into aspiration over time. We therefore hypothesize that spheroid retraction is determined by an interplay between retractile cellular elastic properties and the viscous flow of the spheroid tongue as a cellular collective, each having their own critical pressure governing retraction.
\begin{figure}
\includegraphics[width=8cm]{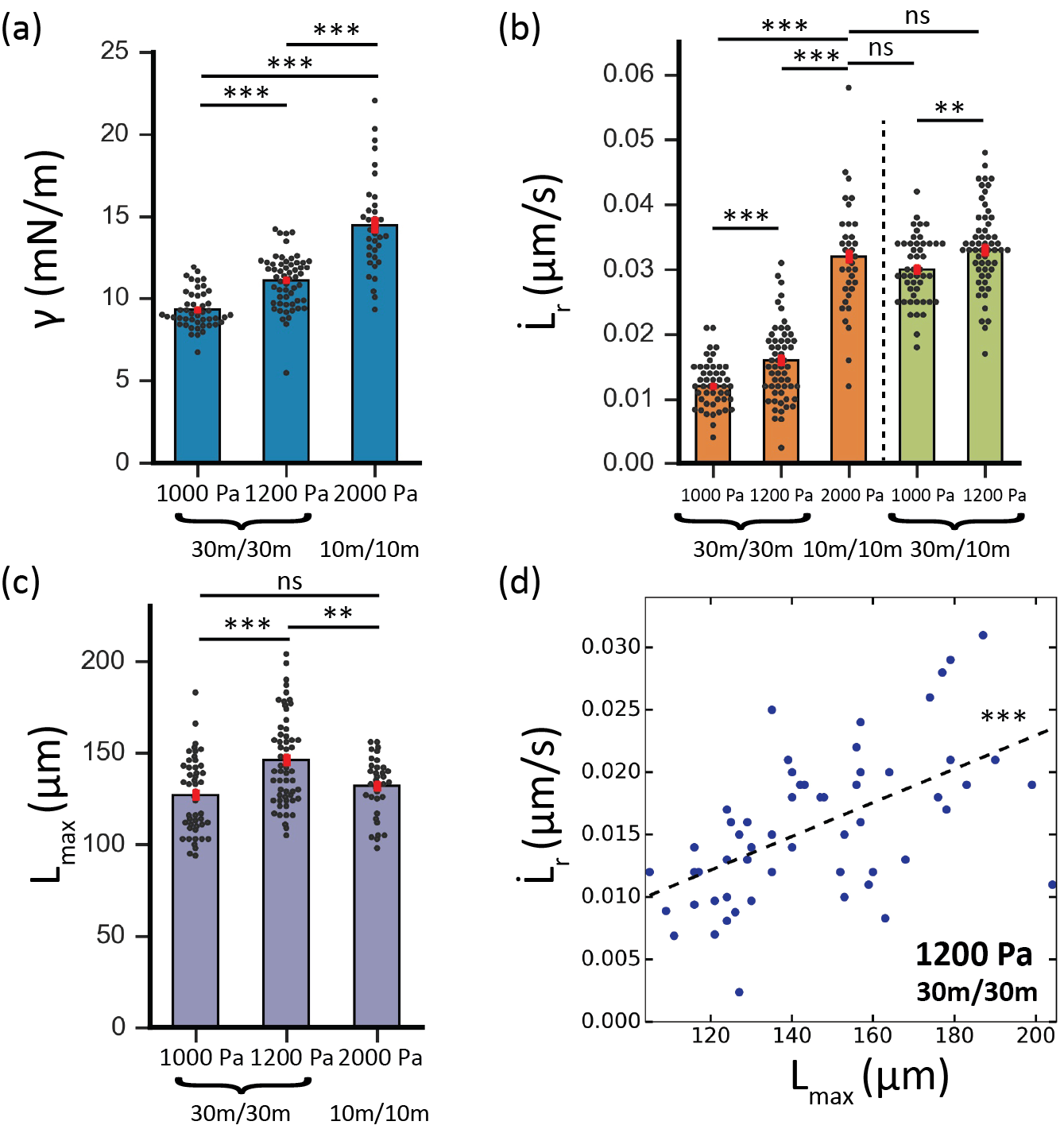}
\caption{\label{Fig2} Tissue relaxation behavior is deformation-dependent. (a-c) Histograms comparing NIH3T3 spheroids aspirated at 1000 Pa (30 min aspiration, 30 min retraction, n = 48), 1200 Pa (30 min aspiration, 30 min retraction, n = 57), and at 2000 Pa (10 min aspiration, 10 min retraction, n = 35). Retraction was performed at 200 Pa. (a) $\gamma$, (b) $\dot{L_\infty^r}$ and (c) $L_{max}$ are compared. For (b), the two green bars on the right of the dotted line depict the derived $\dot{L_\infty^r}$ when only fitting the first 10 min of retraction. (d) $\dot{L_\infty^r}$ plotted versus $L_{max}$ for NIH3T3 spheroids aspirated at 1200 Pa (30 min aspiration, 30 min retraction). **, p $<$ 0.01, ***, p $<$ 0.001 and ns is nonsignificant. Error bars are SEM.}
\end{figure}

To test whether we observe a reinforcement in $\gamma$ for increasing $\Delta P$ as reported in Ref. \cite{Dufour2010}, measurements with NIH3T3 spheroids were conducted using two slightly differing aspiration pressures (1000 Pa and 1200 Pa) and a large pressure of 2000 Pa, now for only 10 min of aspiration followed by 10 min of retraction at 200 Pa as spheroid volumes were not conserved for aspiration times beyond 10 min. At 1000 and 1200 Pa, we observed pulsed contractions or "shivering" in ca. 50\% of the aspiration curves (examples in [Supplementary Fig. 6 \cite{SupMat}]), resembling observations reported with glass MPA \cite{Guevorkian2011}. This shivering effect did not occur at the large pressure of 2000 Pa, where the protrusion flowed smoothly in the constriction. Despite the shivering, aspiration curves could still be fitted with the modified Maxwell model and retraction curves were comparable for spheroids that did or did not shiver during aspiration [Supplementary Table 1 \cite{SupMat}]. The small influence of shivering on $\dot{L_\infty^a}$ did not significantly influence $\dot{L_\infty^r}$ nor $\gamma$ for different conditions, so we included these data. For the three different aspiration pressures, we indeed observed an apparent force-dependent $\gamma$, where the derived surface tension increased for larger $\Delta P$ [Fig. \ref{Fig2}(a)]. Accordingly, the stress relaxation curves demonstrated an increase in retraction flow velocity $\dot{L_\infty^r}$ for larger $\Delta P$ [Fig. \ref{Fig2}(b)], formerly also observed in glass MPA measurements \cite{Dufour2010}. Previously, this was interpreted as the spheroid protrusion actively mechanosensing the magnitude of the aspiration force, causing it to reinforce and retract faster. However, for our measurements $\dot{L_\infty^r}$ depends on the time frame during which the relaxation is investigated. Intriguingly, when fitting only the first 10 min of retraction for the measurement at 1000 Pa, we find that the average $\dot{L_\infty^r}$ at 1000 and 2000 Pa is the same [Fig. \ref{Fig2}(b), right-side]. While the total deformation length $L_{max}$ of the spheroid protrusion at the end of aspiration is understandably larger when aspirating at larger pressures, we find that the same average length has been reached when aspirating for 30 min at 1000 Pa or 10 min at 2000 Pa, indicating a deformation-dependency for retraction [Fig. \ref{Fig2}(c)]. Indeed, we find that $\dot{L_\infty^r}$ is linearly dependent on $L_{max}$, where the further the protrusion has reached after aspiration, the faster it retracts when comparing identical time frames [Fig. \ref{Fig2}(d), Supplementary Fig. 7 (1000 Pa and 2000 Pa) \cite{SupMat}]. In addition, a larger aspiration flow velocity $\dot{L_\infty^a}$ results in a larger $L_{max}$ and thus larger $\dot{L_\infty^r}$ [Supplementary Fig. 8 \cite{SupMat}]. While these trends were observed at 1000 and 1200 Pa, they were not significant for the measurements at 2000 Pa, likely because of the larger standard deviation in $\dot{L_\infty^r}$ at the shorter timescale, alongside the smaller range in $L_{max}$ and the smaller number of data points. Altogether, these measurements show that $\dot{L_\infty^r}$ and the derived $\gamma$ do not solely depend on either the applied force or time frame but directly relate to the product of both, being the resulting length of deformation $L_{max}$. 
\begin{figure}[b]
\includegraphics[width=8cm]{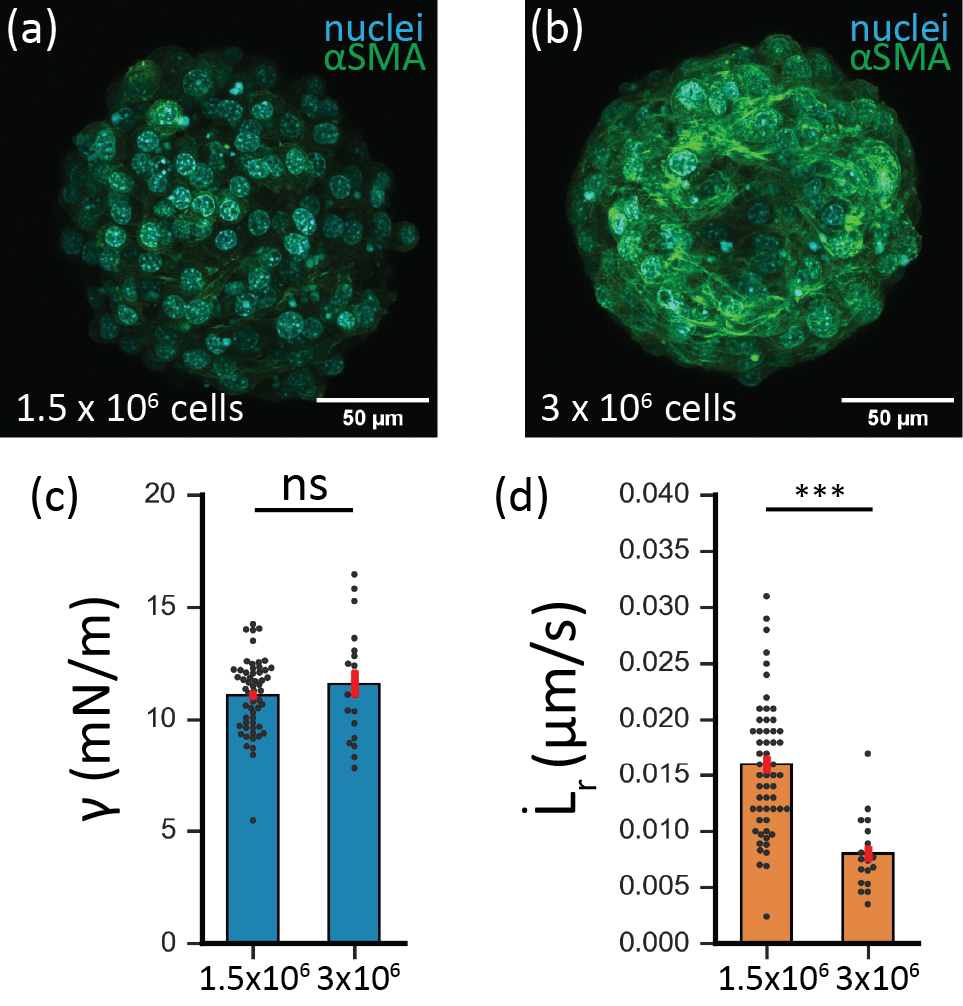}
\caption{\label{Fig3} A larger concentration of \textalpha-SMA does not influence derived tissue surface tension. (a-b) Max intensity confocal fluorescent images of nuclei (cyan) and \textalpha-SMA (green) in NIH3T3 spheroids seeded with (a) 1.5x10\textsuperscript{6} cells/well and (b) 3x10\textsuperscript{6} cells/well. (c-d) Histograms comparing NIH3T3 spheroids aspirated at 1200 Pa at a seeding density of 1.5x10\textsuperscript{6} cells (30 min aspiration, 30 min retraction, n = 57) and a seeding density of 3x10\textsuperscript{6} cells (30 min aspiration, 30 min retraction, n = 19). Retraction was performed at 200 Pa. (c) $\gamma$ and (d) $\dot{L_\infty^r}$ are compared. ***, p $<$ 0.001 and ns is nonsignificant. Error bars in histograms are SEM.}
\end{figure}

What cellular properties govern retraction flow velocity and its deformation-dependency is unclear. We therefore sought to investigate the influence of cell contractility on the retraction flow and spheroid viscoelasticity. Alpha-smooth muscle actin (\textalpha-SMA), the mesenchymal marker and cytoskeletal protein that is incorporated into stress fibers of fibroblasts, upregulates their contractile activity and ability to remodel tissues \cite{Scanlon2013,Hinz2001,Sarrio2008}. We found that increasing the NIH3T3 cell seeding density during fabrication strongly influenced the \textalpha-SMA concentration in our spheroids [Fig. \ref{Fig3}(a-b)]. Western blots analysis showed that doubling the cell seeding density from 1.5x10\textsuperscript{6} (used in Fig. \ref{Fig2}) to 3x10\textsuperscript{6} cells increased the average protein concentration of \textalpha-SMA by a factor of 6 [Supplementary Fig. 9 \cite{SupMat}]. We hypothesize that cells differentiated at higher density, similar to fibroblasts increasing their \textalpha-SMA concentration in response to the cytokine TGF-\textbeta 1 when seeded at a larger cell density in 2D \cite{Doolin2021}. Identical to the previous spheroids, we aspirated the 3x10\textsuperscript{6} cell spheroids at 1200 Pa for 30 min and then let them retract at 200 Pa for 30 min. Our results showed that $\gamma$ did not change, despite the larger concentration of \textalpha-SMA [Fig. \ref{Fig3}(c)]. This was unexpected, as the 3x10\textsuperscript{6} cell spheroids retracted slower [Fig. \ref{Fig3}(d)]. Additionally, they had a significantly lower aspiration rate $\dot{L_\infty^a}$ and reached less far in the constrictions, demonstrating a smaller deformability [Supplementary Fig. 10 \cite{SupMat}]. From this, we calculate that the 3x10\textsuperscript{6} cell spheroids have a higher average viscosity of $\eta$ = 150 $\pm$ 9 kPa.s in comparison to the 1.5x10\textsuperscript{6} cell spheroids with $\eta$ = 68 $\pm$ 3 kPa.s [details in Supplemental Material \cite{SupMat}], explaining their different deformability at the same $\gamma$. In contrast, previous measurements with parallel-plate tensiometry on spheroids composed of different cell lines showed a linear correlation between bulk viscosity and surface tension \cite{Yu2018}. However, our findings on a time-dependent retraction raise the question whether the viscosity and/or $\Delta P_c$ remain constant during MPA retraction. If this assumption would prove to be incorrect, disentangling $\eta$ and $\gamma$ when interpreting the measured $\dot{L_\infty^a}$ and $\dot{L_\infty^r}$ becomes very difficult, as we are left with two separate responses (aspiration and retraction) each with two unknown variables. 

\begin{figure}
\includegraphics[width=8.2cm]{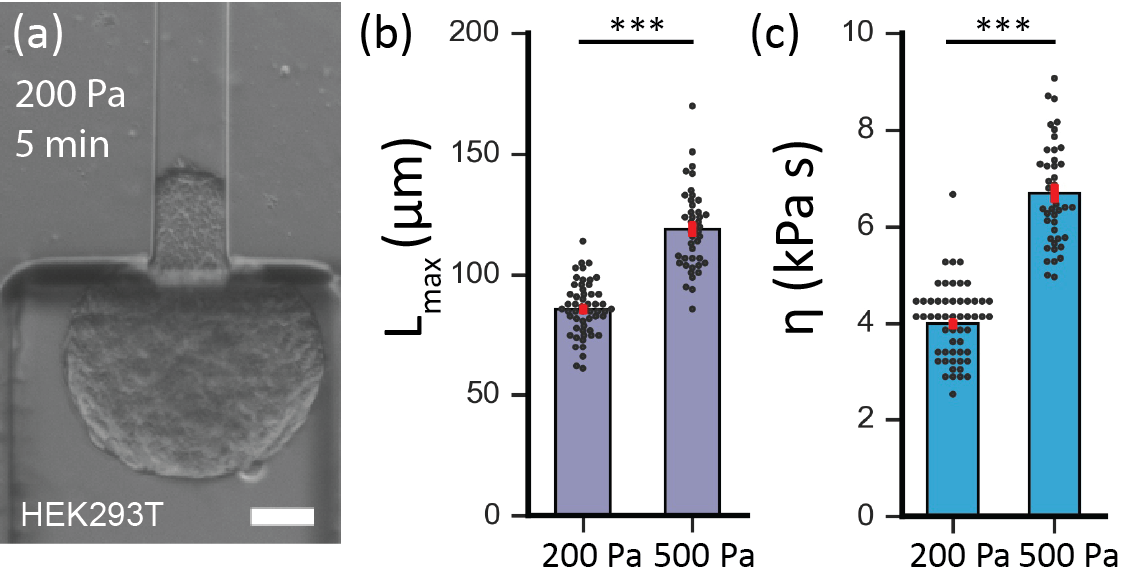}
\caption{\label{Fig4} For soft HEK293T cell spheroids, viscosity is pressure-dependent. (a) Brightfield image of a HEK293T cell spheroid after 5 min of aspiration at 200 Pa. Scale bar 50 \textmu m. (b-c) Histograms comparing HEK293T spheroids aspirated at 200 Pa (5 min aspiration, 5 min retraction, n = 54) and at 500 Pa (5 min aspiration, 5 min retraction, n = 43). Retraction was performed at 50 Pa. (b) $L_{max}$ and (c) $\eta$ are compared. ***, p $<$ 0.001. Error bars are SEM.}
\end{figure}

To untangle $\eta$ and $\Delta P_c$, we aspirated spheroids without the influence of $\Delta P_c$, using human embryonic kidney (HEK293T) cell spheroids with a very low surface tension [Fig. \ref{Fig4}(a)]. In order to keep spheroid volume constant, we aspirated the HEK293T cell spheroids for 5 min at 200 Pa and 500 Pa. We then monitored retraction for 5 min at a remaining minor pressure of 50 Pa. After 200 Pa aspiration, spheroid tongues retracted elastically and then started aspirating again, indicating that their $\Delta P_c$ was indeed minimal (below 50 Pa) [Supplementary Fig. 11 \cite{SupMat}]. Upon 500 Pa aspiration, spheroid tongues deformed further into the constrictions [Fig. \ref{Fig4}(b)] and now displayed a linear viscous retraction over time, resulting in an average $\gamma$ = 1.9 $\pm$ 0.1 mN/m, consistent with a deformation-dependent retraction. Importantly, at these low pressures we find a pressure-dependent viscosity [Fig. \ref{Fig4}(c)]. This contrasts with our NIH3T3 measurements, where the viscosity was pressure-independent [Supplementary Fig. 12(a) \cite{SupMat}], similar to the previous study by Guevorkian \textit{et al.} with S180 cell spheroids \cite{Dufour2010}. In our own previous study with a slightly different microfluidic design (aspiration pockets were rounded instead of rectangular) \cite{Boot2022}, we aspirated HEK293T spheroids at 500 and 700 Pa and also showed their viscosity to be pressure-independent. We believe that the discrepancy with our new data can be explained by the smaller pressure range (factor 1.4, compared to a factor 2.5 in the current work) and the larger standard deviation in $\dot{L_\infty^a}$ for our previous microfluidic device. Our new measurements clearly display a significant increase in $\eta$ when raising the pressure, from $\eta$ = 4.0 $\pm$ 0.1 kPa.s at 200 Pa to $\eta$ = 6.7 $\pm$ 0.2 kPa.s at 500 Pa [Fig. \ref{Fig4}(c)]. For 200 Pa, we calculated $\eta$ by assuming $\Delta P_c$ = 0 Pa, thus giving an upper bound for $\eta$. For 500 Pa, we calculated $\eta$ via our derived $\Delta P_c$ from $\dot{L_\infty^a}$ and $\dot{L_\infty^r}$. Interestingly, the proportional change in the initial elastic deformation $\delta_a$ during aspiration was larger than the change in $\dot{L_\infty^a}$ between the two pressures [Supplementary Fig. 13 \cite{SupMat}], suggesting that the increase in force differently affects the elastic deformation and the viscous flow of cells. Moreover, we do find a pressure-dependent $\eta$ for our NIH3T3 spheroid measurements when we reanalyze the data, neglecting the increase in $\dot{L_\infty^r}$ and using that $\Delta P_c$ is the same at 1000 Pa and 1200 Pa [Supplementary Fig. 12(a) \cite{SupMat}]. Similar to the HEK293T measurements, the increase in force induces a larger proportional change in $\delta_a$ than for $\dot{L_\infty^a}$ [Supplementary Fig. 12(b-c) \cite{SupMat}]. This implies that another possible framework exists besides tissue reinforcement, in which not surface tension but spheroid viscosity is pressure-dependent for MPA measurements. In that case, either $\eta$ or $\Delta P_c$ is different between aspiration and retraction, as the flow velocity is deformation-dependent during retraction.

In this study, we have measured a force-dependent spheroid surface tension $\gamma$ coupled to an increased viscous flow rate $\dot{L_\infty^r}$ at larger deformations. The reinforcement of $\gamma$ has previously been explained by an active response of cells to mechanical forces, involving cytoskeletal remodeling potentially due to stress fiber polymerization by myosin II motors, stretch-activated membrane channels or the clustering of cadherins \cite{Dufour2010,Janmey2004,Martino2018,Sbrana2008,Delanoe2004,Ingber2006,Kaunas2011}. Next to this, spheroid surface tension has also previously been coupled to the size of spheroids \cite{Yousafzai2022}, where spheroids in the size range of 160-360 \textmu m in diameter displayed a smaller $\gamma$ as the size increased. Unfortunately, the size range that our microfluidic device can aspirate was too small to reproduce this effect [Supplementary Fig. 14 \cite{SupMat}]. However, our results have generated new insights in the viscoelastic behavior of spheroids during MPA and the interpretation of $\gamma$ thanks to the large amount of data we could obtain by high-throughput microfluidic aspiration. We therefore propose a different framework than tissue reinforcement to interpret viscoelastic spheroid MPA data, in which $\dot{L_\infty^r}$ is governed by the total deformation $L_{max}$, making $\eta$ and/or $\Delta P_c$ different between aspiration and retraction. 

How to distinguish tissue surface tension and viscosity from each other during aspiration measurements, and how to identify whether $\gamma$, $\eta$ or both are pressure-dependent during MPA remains an open question. Previous studies have shown how the liquid-like properties of cellular tissues are determined by tissue flow via cells rearranging and slipping past each other \cite{Foty2004,Grosser2021,Kosheleva2020,Pawlizak2015,Marmottant2009,David2014}. In an experiment with mCherry-transfected NIH3T3 cells and a microfluidic device modified to allow detailed imaging in the constriction channel, we did not observe cells slipping past each other, demonstrating that the viscous flow was unlikely to be governed by cell rearrangements [see Supplementary Movies 1 and 2, and Supplemental Material for the device modifications \cite{SupMat}]. We hypothesize that the deformation-dependent viscoelasticity can be explained by the number of individual cells that have been aspirated into the constriction channel. If each cell has its own relaxation rate, then a spheroid protrusion with more cells in series will have a larger $\dot{L_\infty^r}$, being the sum of all these individual cellular retraction rates. Yet, how the tissues' effective viscosity is precisely governed at the cellular level, and how different cytoskeletal elastic and viscous components work over different timescales to govern $\Delta P_c$ remains to be examined. Overall, we show that spheroid viscoelastic behavior is pressure- and deformation-dependent for MPA, challenging the assumption that both $\eta$ and $\Delta P_c$ are identical during aspiration and retraction. 

\begin{acknowledgments}
R.C.B. and P.E.B. gratefully acknowledge funding from the European Research Council (ERC) under the European Union’s Horizon 2020 research and innovation program (grant agreement no. 819424). P.M. and P.E.B. gratefully acknowledge funding from the Delft Health Technology grant. G.H.K. gratefully acknowledges funding from the VICI project \textit{How cytoskeletal teamwork makes cells strong} (project number VI.C.182.004) and from an OCENW.GROOT.20t9.O22 grant ('The Active Matter Physics of Collective
Metastasis'), both financed by the Dutch Research Council (NWO). The authors thank Peter ten Dijke's laboratory (at LUMC) for technical support in the cell transfection, and Karine Guevorkian and Timon Idema for helpful discussions.
\end{acknowledgments}

\bibliography{apssamp}% Produces the bibliography via BibTeX.

\end{document}